\documentclass[letterpaper,prl,twocolumn]{revtex4}
\usepackage{graphicx}
\usepackage{dcolumn}
\usepackage{bm}
\usepackage{xcolor}
\usepackage{soul}
\usepackage[english]{babel}
\usepackage{epsfig}
\usepackage{amsmath}
\usepackage{amsfonts}
\usepackage{amssymb}
\usepackage{float}
\usepackage{mathrsfs}
\usepackage{mathtools}

\graphicspath{{eps/}}

\begin{document}

\title{Direct scattering transform: catch soliton if you can} 

\author{Andrey Gelash$^{1,2}$}\email{agelash@gmail.com}
\author{Rustam Mullyadzhanov$^{3,4}$}\email{rustammul@gmail.com}

\affiliation{$^{1}$Institute of Automation and Electrometry SB RAS, Novosibirsk 630090, Russia}
\affiliation{$^{2}$Skolkovo Institute of Science and Technology, Moscow 121205, Russia}
\affiliation{$^{3}$Institute of Thermophysics SB RAS, Novosibirsk 630090, Russia}
\affiliation{$^{4}$Novosibirsk State University, Novosibirsk 630090, Russia}

\begin{abstract}

Direct scattering transform of nonlinear wave fields with solitons may lead to anomalous numerical errors of soliton phase and position parameters.
With the focusing one-dimensional nonlinear Schr\"odinger equation serving as a model, we investigate this fundamental issue theoretically.
Using the dressing method we find the landscape of soliton scattering coefficients in the plane of the complex spectral parameter for multi-soliton wave fields truncated within a finite domain, allowing us to capture the nature of particular numerical errors.
They depend on the size of the computational domain $L$ leading to a counterintuitive exponential divergence when increasing $L$ in the presence of a small uncertainty in soliton eigenvalues.
In contrast to classical textbooks, we reveal how one of the scattering coefficients loses its analytical properties due to the lack of the wave field compact support in case of $L \to \infty$.
Finally, we demonstrate that despite this inherit direct scattering transform feature, the wave fields of arbitrary complexity can be reliably analysed.

\end{abstract}

\maketitle

{\it Introduction}. -- 
Since the 1970s we have observed an impressive progress in nonlinear mathematical physics stimulated by the discovery of the complete integrability of some nonlinear partial differential equations \cite{NovikovBook1984, AblowitzBook1981, NewellBook1985, OsborneBook2010}.
Among them are the one-dimensional Korteweg--de Vries (KdV) \cite{GGKM1967} and nonlinear Schr\"odinger (NLSE) \cite{Zakharov1972} equations serving as the fundamental nonlinear wave models and appearing in different areas of physics.
This breaktrough has taken place due to the development of the {\it inverse scattering transform} (IST) \cite{NovikovBook1984, AblowitzBook1981} allowing one to solve the initial-value problem in terms of the nonlinear harmonics decomposition representing the {\it scattering data} or the {\it IST spectrum}.
The spectrum can be found using the {\it direct scattering transform} (DST) leading to the full knowledge of the nonlinear wave field evolution governed by the integrable differential equation \cite{NovikovBook1984, AblowitzBook1981}.
After several decades of analytical studies of integrable equations, the rapid growth of interest to describe
arbitrary shaped, noisy and even random nonlinear wave fields has promoted the need in accurate numerical methods for the DST.
The Boffetta--Osborne method \cite{BofOsb1992} represents the first numerical realization of the DST followed by a sequence of further improvements and alternatives \cite{Burtsev1998, Frumin2015, vasylchenkova2018contour, Fedoruk2019, Aref2019}.
These advancements have made the DST an essential scientific tool with a wide range of theoretical and experimental applications
\cite{OsborneBook2010,prilepsky2014nonlinear,Tavakkolnia:17,Turitsyn2017nonlinear,slunyaev2018analysis,Randoux2018nonlinear}.
The remarkable ability of the DST to identify and characterise solitons representing the coherent structures in nonlinearly interacting wave fields provides fundamental information about the origin of various physical effects \cite{derevyanko2012nonlinear, Gelash2018} and can be fruitfully used in practical applications such as optical telecommunication systems \cite{turitsyn2008soliton,yousefi2014information, dong2015nonlinear, Frumin2017}.
A significant amount of work has been devoted to understand the distribution of soliton amplitudes and velocities (eigenvalues) \cite{bronski1996semiclassical,klaus2002purely} with particular focus on its role in propagation of ocean waves \cite{osborne1991soliton, yousefi2014information, slunyaev2018analysis} and optical pulses \cite{derevyanko2012nonlinear, braud2016solitonization}.
Although the soliton eigenvalues can be found using many variations of the numerical DST \cite{YangBook, Vasylchenkova2019}, an accurate identification of soliton phase and position parameters represented by the so-called {\it norming constants} is still a challenging problem.
So far the existing approaches for finding both eigenvalues and norming constants have demonstrated success only for a relatively simple wave fields containing up to five solitons \cite{BofOsb1992, Aref2019, Vasylchenkova2019}.
The recent development of high-order numerical schemes has allowed to process large multi-soliton wave fields with particular examples of 128 solitons, revealing several types of numerical instabilities including the non-trivial behaviour of norming constants \cite{Mullyadzhanov2019}.
The first class of the instabilities is a result of accumulation of discretization errors during the scattering through a large wave field, which can be efficiently resolved by high-order schemes \cite{Mullyadzhanov2019}.
The second class is related to round-off errors during the computation of the scattering coefficients, which can be fixed by high-precision arithmetics similar to the IST procedure \cite{Gelash2018}.
The third class represents {\it anomalous} errors for the norming constants, when the eigenvalues are computed without sufficient accuracy, requiring high-precision arithmetics for eigenvalue identification.
In this work we theoretically reveal the nature of these anomalous errors within the DST framework making soliton phase and position characteristics extremely elusive.
As a model we consider the focusing NLSE for a complex wave field $\psi(t, x)$, which 
in the non-dimensional form is as follows:
\begin{eqnarray}
\label{eqNLS}    i \psi_t + \frac12 \psi_{xx} + |\psi|^2 \psi = 0,
\end{eqnarray}
where $t$ and $x$ are the time and spatial coordinate.
The soliton parameters represent the discrete part of the scattering data,
which can be found via the so-called {\it scattering coefficients}.
Instead of standard scattering coefficients defined on an infinite line, we use their ``truncated'' analogues on a finite interval leading to the sensible theoretical analysis of the error sources within the DST.
In particular, employing the dressing method, we find the landscape of the scattering coefficients
in the complex plane of the spectral parameter for a multi-soliton wave field and track the behaviour of numerical errors for different configurations of scattering data leading to specific recipes on error reduction.

We show that these errors for norming constants depend on the size of the computational domain $L$ leading to a counterintuitive exponential divergence when increasing $L$ in the presence of a small uncertainty in soliton eigenvalues.
In contrast to classical textbooks, we study how one of the scattering coefficients loses its analytical properties due to the lack of the wave field compact support in case of $L \to \infty$.
Finally, we demonstrate that despite this inherit DST feature, the wave fields of arbitrary complexity can be reliably analysed.

{\it The direct scattering transform for the NLSE}. --
The IST theory establishes a link between the focusing NLSE, see Eq. (\ref{eqNLS}), and the following auxiliary Zakharov--Shabat (ZS) linear system for the two-component vector wave function $\mathbf{\Phi} (t, x, \zeta) = (\phi_1,\phi_2)$ \cite{Zakharov1972}:
\begin{eqnarray}
\mathbf{\Phi}_{x} = \mathbf{\widehat{Q}}(\psi) \mathbf{\Phi}, \;\;\; \mathbf{\widehat{Q}} = 
\begin{pmatrix}
-i \zeta & \psi \\ -\psi^* & i \zeta
\end{pmatrix},
\label{ZSsystem}
\end{eqnarray}
where $\zeta = \xi + i \eta$ is a spectral parameter, the star denotes the complex-conjugate value.
The ZS system features a spectrum composed of continuous and discrete parts with the latter located on the complex plane.
As typically done, we consider the upper half of the complex plane with $\eta \geqslant 0$ since $\zeta^* = \xi - i \eta$ correspond to the same class of NLSE solutions.
The continuous spectrum occupies only the real axis $\xi \in \mathbb{R}$, while the discrete part (eigenvalues) is represented by the complex points with its total number equal to $N$.
The scattering data of the potential $\psi (t, x)$ is traditionally introduced using a solution of the ZS system with $\zeta=\xi$ and the following asymptotics at infinity:
\begin{equation}
\label{ScatteringProblem}
\lim_{x \to -\infty}
\mathbf{\Phi} = 
\begin{pmatrix}
e^{-i \xi x} \\ 0
\end{pmatrix}
,\,\,\,\,\,
\lim_{x \to \infty}
\mathbf{\Phi} = 
\begin{pmatrix}
a(\xi) e^{-i \xi x} 
\\ b(\xi) e^{i \xi x}
\end{pmatrix},
\end{equation}
where $a(\xi)$ and $b(\xi)$ are the scattering coefficients.

\begin{figure} 
    \includegraphics[width=0.5\textwidth]{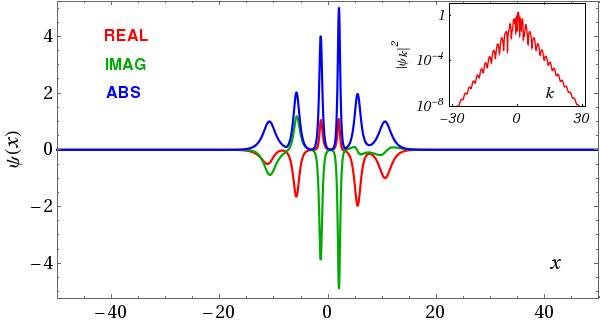}%
    \caption{
        Typical multi-soliton wave field used for demonstration of the DST numerical errors with $N = 6$. 
        The 6-$SS$ solution is obtained using Eqs. (\ref{N-SS}) and (\ref{N-SS details}).
        See Fig. \ref{fig_2} on the spectral content of this solution.
    }
    \label{fig_1}
\end{figure}

The first coefficient has an analytic continuation $a(\zeta)$ to the $\zeta$-plane with zeros at the discrete eigenvalues $\zeta_k$ with $k = 1,...,N$.
The second coefficient $b(\xi)$ is defined on the real axis and at the eigenvalue points $\zeta_k$ with $b(\zeta_k) = b_k$.
It is important to emphasize that $b(\xi)$ can be analytically continued to the $\zeta$-plane only when the potential $\psi(x)$ has {\it compact support}, i.e. if $\psi = 0$ outside of a compact set on the $x$-line \cite{faddeev2007hamiltonian}.
As mentioned earlier, the total scattering data represents the discrete $\{\zeta_k, \rho_k \}$ and continuous $\{ r \}$ spectrum:
\begin{equation}
\label{ScatteringData}
a(\zeta_k)=0,
\,\,\,\,\,\,
\rho_k = \frac{b_k}{a'(\zeta)} \bigg |_{\zeta=\zeta_k};
\,\,\,\,\,\,
r(\xi) = \frac{b(\xi)}{a(\xi)},
\end{equation}
where $\rho_k$ is the complex-valued norming constants associated with $\zeta_k$ and $r(\xi)$ is the reflection coefficient.
Each discrete eigenvalue $\zeta_k = \xi_k + i\eta_k$ corresponds to a soliton in the wave field with the amplitude $A_k=2\eta_k$ and group velocity $V_k=2\xi_k$, while $r(\xi)$ describes nonlinear dispersive waves.
In case of $r(\xi)=0$, i.e. the dispersive waves are absent, the wave field corresponds to $N$-soliton solution $\psi_{NSS}(x)$ which can be reconstructed analytically employing the scattering data (\ref{ScatteringData}) \cite{NovikovBook1984} (see also \cite{Mullyadzhanov2019}):
\begin{equation}
\label{N-SS}
\psi_{NSS}(x) = 
-2 i \rho_k e^{i\zeta_k x} \,
[(\mathbf{E} + \mathbf{M}^*\mathbf{M})^{-1}]_{k,j} \, e^{i\zeta_j x}.
\end{equation}
Here $\mathbf{E}$ is the $N\times N$ unity matrix and the elements:
\begin{eqnarray}
\label{N-SS details}
\mathbf{M}_{k,j} = i\rho_j (\zeta^*_k - \zeta_j)^{-1} e^{-i (\zeta^*_k - \zeta_j) x}.
\end{eqnarray}
The norming constant can be conveniently parametrized as follows:
\begin{equation}
\label{rhok_param}
\rho_k = -i A_k e^{A_k x_{0_k} - i\theta_k},
\end{equation}
where two real-valued parameters $x_{0_k}$ and $\theta_k$ describe the position in space and phase of the corresponding soliton \cite{LambBook1980}.
Below we use an example shown in Fig. \ref{fig_1} demonstrating a typical multi-soliton wavepacket with $N = 6$ (6-$SS$) constructed with the help of the above formulas with a random distribution of $\{ \theta_k \}$ and chosen set of $\{ x_{0_k} \}$.

For $\psi_{NSS}(x)$ the first scattering coefficient is known in a closed form in the $\zeta$-plane \cite{NovikovBook1984}:  
\begin{equation}
\label{a_expr_NSS}
a_{N}(\zeta) = \prod_{k=1}^{N} \frac{\zeta-\zeta_k}{\zeta-\zeta^*_k}\, ,
\end{equation}
while the second scattering coefficient $b_{N}(\xi)$ cannot be analytically continued to $\zeta$-plane due to infinite exponentially decaying tails of $\psi_{NSS}(x)$ contradicting to the necessary compact support property.
{\it Truncation of the wave field}. -- 
We begin our theoretical analysis of the DST introducing a finite domain 
$[-L, L]$, where the wave field $\psi(x)$ is well localized.
The truncation of $\psi$ guarantees the compact support allowing to define $b_{tr}(\zeta)$ in the  complex plane, where the subscript `$tr$' highlights that the corresponding $\psi$ is non-zero only inside the domain $[-L, L]$.
We introduce a wave function $\mathbf{\Phi}_{tr}$ with shifted boundary conditions from $x \to \pm \infty$, see Eq. (\ref{ScatteringProblem}), to $x = \pm L$:
\begin{equation}
\label{ScatteringProblemTrunct}
\mathbf{\Phi}_{tr}(-L) = 
\begin{pmatrix}
e^{-i \zeta L} \\ 0
\end{pmatrix}
,\,\,\,\,\,
\mathbf{\Phi}_{tr}(L) = 
\begin{pmatrix}
a_{tr}(\zeta) e^{-i \zeta L} 
\\ b_{tr}(\zeta) e^{i \zeta L}
\end{pmatrix},
\end{equation}
while with the help of $a_{tr}(\zeta)$ and $b_{tr}(\zeta)$ we define
\begin{equation}
\label{rho(zeta)}
\rho_{tr}(\zeta) = \frac{b_{tr}(\zeta)}{a'_{tr}(\zeta)}. 
\end{equation}

\begin{figure} 
    \centering
    \includegraphics[width=0.25\textwidth]{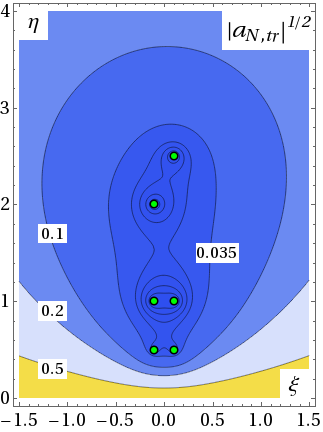}%
    \includegraphics[width=0.25\textwidth]{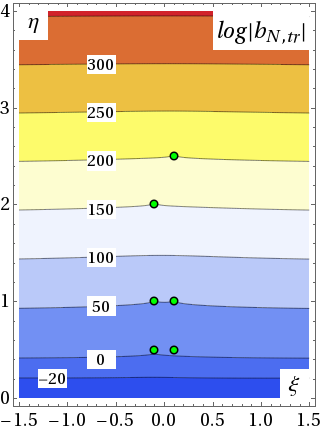}%
    \caption{
        Behaviour of the scattering coefficients in $\zeta$-plane according to Eqns. (\ref{a_IST_N_conv}) and (\ref{b_IST_N_conv}) with $L = 50$. %
        Green dots show $\{ \zeta_k \}$ corresponding to the wave packet demonstrated in Fig. \ref{fig_1}.
    }
    \label{fig_2}
\end{figure}

Our key result is the theoretical derivation of $a_{N,tr}(\zeta)$ and $b_{N,tr}(\zeta)$ corresponding to $N$-soliton potentials in the $\zeta$-plane. 
Using the dressing method to construct solutions of ZS problem for $\psi_{NSS}$ and assuming large enough $L$, we obtain the following expressions (see Supplementary materials for a detailed derivation):
\begin{eqnarray}
\label{a_IST_N_conv}
a_{N,tr}(\zeta) &=& a_N(\zeta) +  o_N, \\
\label{b_IST_N_conv}
b_{N,tr}(\zeta) &=& a_N(\zeta)\sum_{k = 1}^N
\frac{\rho_k e^{-2 i (\zeta - \zeta_k) L}}{\zeta - \zeta_k} + o_N.
\end{eqnarray}
It is convenient to extract $a_N(\zeta)$ in Eq. (\ref{b_IST_N_conv}), although it cancels out the denominator $(\zeta - \zeta_k)^{-1}$ for all $k$ when the full expression (\ref{a_expr_NSS}) is explicitly employed.
A new notation is used to shorten the presentation:
\begin{equation}
    o_N = p(\zeta) o(e^{- 2\eta_{min} L}),
\end{equation}
which is based on a ``little-$o$'' \cite{hardy1979introduction}, a rational function $p(\zeta)$ and the expression $\eta_{min} = \text{min} [ \eta_1,..,\eta_N ]$ representing the minimum value among the considered set of $\eta$.
Fig. \ref{fig_2} shows the typical behaviour of $a_{N, tr}(\zeta)$ around the eigenvalues and stiff exponential growth of $b_{N, tr}(\zeta)$ for the 6-$SS$ solution presented in Fig. \ref{fig_1}.
We should stress that both formulas (\ref{a_IST_N_conv}) and (\ref{b_IST_N_conv}) are verified numerically in the $\zeta$-plane, see Supplementary materials.
However, concerning the accuracy of $b_{N, tr}$ close to the real axis a small deviation can still arise due to the fact that the leading order term and $o_N$ become of the same order.
This region of interest has recently been addressed \cite{aref2019nonlinear}.

The expression (\ref{b_IST_N_conv}) covers a fundamental issue on the analytical properties of $b_{N}(\zeta)$ in the complex plane leading to the following result:
\begin{eqnarray}
\label{bNlim}
b_{N}(\zeta) = \lim_{L \to \infty} b_{N,tr}(\zeta) =
   \begin{dcases}
     b_k, \,\,\,\,\,\,\,\,\,\,\,\,\,\,\,\,\,\,\, \zeta=\zeta_k, \\
     \infty, \,\,\,\,\,\, \eta > \eta_{min}; \,\,\, \zeta\ne \zeta_k, \\
     b_{min}(\zeta), \,\,\,\,\, \eta = \eta_{min}, \\
     0, \,\,\,\,\,\,\,\,\,\,\,\,\,\,\,\,\,\,\,\,\, \eta < \eta_{min},
   \end{dcases}
\end{eqnarray}
where $b_{min}(\zeta) = a_N(\zeta) \rho_{min} e^{-2 i (\zeta - \zeta_{min}) L}/(\zeta - \zeta_{min})$ with the subscript `$min$' corresponding to the soliton with the minimal $\eta$ as defined above.
Thus, the second scattering coefficient $b_N(\zeta)$ is analytic only inside the band $0 \le \eta < \eta_{min}$.

{\it Anomalous errors}. -- 
Ideally, when $L$ tends to infinity, one expects to end up with the exact formulation of the problem, see Eq. (\ref{ScatteringProblem}).
However, one of the counter-intuitive results of the present work boils down to the fact that increasing $L$ leads to a number of numerical difficulties when trying to determine norming constants.
Indeed, expressions (\ref{rho(zeta)}) and (\ref{b_IST_N_conv}) show that at large $L$ even a small deviation $\delta \zeta_k$ of the corresponding computed eigenvalue $\zeta_k$ can lead to large errors for a norming constant $\rho_k$.
In order to show that, we expand the second scattering coefficient (\ref{b_IST_N_conv}) in the vicinity of $\zeta_k$:
\begin{eqnarray}
\label{b_IST_N_lamba_k}
&& b_{N,tr}(\zeta_k + \delta\zeta_k) \approx \underbrace{\frac{\rho_k}{\zeta_k - \zeta_k^*} \prod_{j \ne k}^N \frac{\zeta_k - \zeta_j}{\zeta_k - \zeta_j^*}}_{\text{I term}}  +  \\
&& \underbrace{\delta \zeta_k \Big(  \sum_{l \ne k}^N \frac{\rho_l e^{- 2 i (\zeta_k - \zeta_l) L}}{(\zeta_k - \zeta_l) (\zeta_k - \zeta_k^*)} \prod_{j \ne k}^N \frac{\zeta_k - \zeta_j}{\zeta_k - \zeta_j^*}  \Big)}_{\text{II term}}.     \nonumber
\end{eqnarray}
According to the definition (\ref{rho(zeta)}), the deviation in the norming coefficient $\rho_{N, tr}(\zeta_k + \delta \zeta_k)$ may be exponentially large, caused by the second term in (\ref{b_IST_N_lamba_k}) for $b_{N, tr}$, while $a'_{N, tr}(\zeta_k + \delta \zeta_k)$ does not have this problem.
For each soliton the value of the error, being a function of $L$, can be estimated using the largest exponent of the term II, see (\ref{b_IST_N_lamba_k}):
\begin{equation}
\label{errors_rate}
    \text{error}[\rho_k](L) \sim \text{term II} \sim e^{2 (\eta_k - \eta_{min}) L}.
\end{equation}
The error becomes critical with the increase of $L$ when both terms are of the same order in the expression (\ref{b_IST_N_lamba_k}):
\begin{equation}
\label{delta_lambda_N}
\text{term I} \sim \text{term II}.
\end{equation}
To get a feeling on the order of the deviation $\delta \zeta$ leading to the condition (\ref{delta_lambda_N}), one can inspect the results for a simple two-soliton case:
\begin{eqnarray}
\label{delta_lambda_12}
\delta \zeta^{cr}_1 &\sim& \frac{\rho_1}{\rho_2} (\zeta_1 - \zeta_2) e^{- 2 i (\xi_2 - \xi_1) L} e^{2 (\eta_2 - \eta_1) L},
\\
\label{delta_lambda_21}
\delta \zeta^{cr}_2 &\sim& \frac{\rho_2}{\rho_1} (\zeta_2 - \zeta_1) e^{- 2 i (\xi_1 - \xi_2) L} e^{2 (\eta_1 - \eta_2) L},
\end{eqnarray}
where the subscript `$cr$' denotes the critical value.
Assuming $\eta_2 > \eta_1$ without loss of generality, we obtain an exponential divergence of $\delta \zeta^{cr}_1$ with $L$, while $\delta \zeta^{cr}_2$ on the contrary tends to zero.
This fact means that in order to reduce the error when computing $b_{N, tr}$, one has to guarantee that the eigenvalue is computed with the appropriate accuracy, being demanding for $\delta \zeta^{cr}_2 \to 0$.
The particular number of necessary digits can be estimated from the eigenvalue difference and the value of $L$, although it is obvious that the required precision in majority of cases is more demanding than $10^{-16}$ corresponding to the standard machine precision.
At the same time $\delta \zeta^{cr}_1 \to \infty$ as $L \to \infty$ suggesting that the soliton with the smallest eigenvalue does not suffer from this instability.
Note that $\delta \zeta^{cr}_{1,2}$ in (\ref{delta_lambda_12}), (\ref{delta_lambda_21}) also contain the ratio of norming constants which exponentially depend on the soliton positions $x_{0_{1,2}}$, see the parametrization (\ref{rhok_param}), which can also be an additional stiff condition.
This picture stays the same when more solitons in the wavepacket are considered as is shown below with a numerical example.

{\it Numerical discretization and examples}. -- 
A practical implementation of the DST implies the influence of the numerical discretization on the obtained analytic results.
The eigenvalue condition $a(\zeta_k) = 0$, see Eq. (\ref{ScatteringData}), transforms into:
\begin{equation}
\label{a_num}
a^{num}(\zeta^{num}_{k})=0,
\end{equation}
where the superscript `$num$' denotes the numerical (discretized) counterpart of the initial problem.
Note that this kind of errors should not be confused with ones caused by the domain truncation.
As was shown for $N$-soliton potentials, the relation $a_{N,tr} = a_{N}$ is accurate within exponentially small terms $o_N$ for large $L$, see Eq. (\ref{a_IST_N_conv}), meaning that the discrete spectrum $\{ \zeta_k \}$ does not change within $o_N$ after truncation.
We provide numerical results proving this fact, see Supplementary materials.

We argue that expressions for scattering coefficients (\ref{a_IST_N_conv}) and (\ref{b_IST_N_conv}) for the truncated $\psi_{NSS}(x)$ preserve the same structure after discretization, although exact values $\{ \zeta_k \}$ slightly shift to $\{ \zeta^{num}_k \}$.
This hypothesis is strongly supported by the following numerical results presented below and Supplementary materials.
In the end we study the features of $b^{num}_{N, tr}$ showing that the numerical results are in good agreement with expression (\ref{b_IST_N_conv}), although additional sources of errors are involved but turn out to be irrelevant.

\begin{figure}
    \centering
    \includegraphics[width=0.99\linewidth]{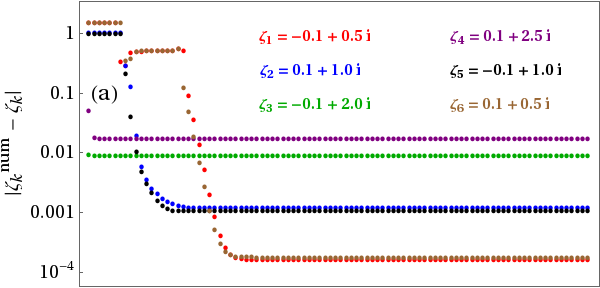}%
    \\
    \includegraphics[width=0.99\linewidth]{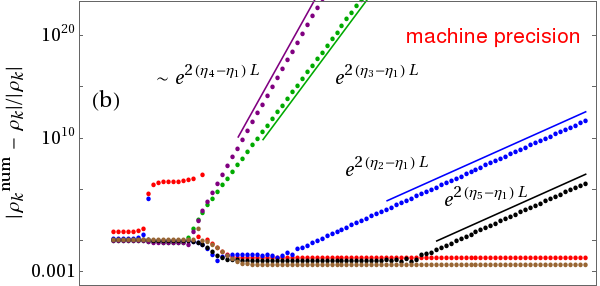}%
    \\
    \includegraphics[width=0.99\linewidth]{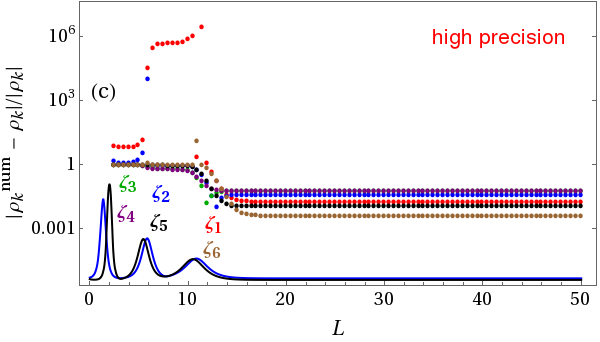}%
    \caption{
        Influence of the size of the numerical domain on the errors of $\{ \zeta_k^{num} \}$ and $\{ \rho_k^{num} \}$ for $6$-$SS$ presented above, see Figs. \ref{fig_1}, \ref{fig_2}.
        (a) Absolute errors for soliton eigenvalues. (b) Relative errors for soliton norming constants computed using $\{ \zeta_k^{num} \}$ obtained with a standard machine precision and (c) high-precision arithmetics. In the bottom we show a schematics for the right ($x > 0$, black) and left ($x < 0$, blue) part of the wave field as in Fig. \ref{fig_1}. 
    }
    \label{fig_3}
\end{figure}
We perform the numerical DST of 6-$SS$ solution presented above as an example, see Figs. \ref{fig_1} and \ref{fig_2}, varying the width of the domain $[-L, L]$.
Fig. \ref{fig_3} demonstrates the influence of $L$ on the errors of the computed discrete spectrum $\{ \zeta^{num}_k, \rho^{num}_k \}$ compared to exact values.
For that we solved the ZS system (\ref{ZSsystem}) using the standard second-order accuracy Boffetta--Osborne method \cite{BofOsb1992} keeping the discretization step constant for all the cases.
In Supplementary materials we verify that high-order methods \cite{Mullyadzhanov2019} demonstrate similar results.
When $L$ is large enough, $\{ \zeta^{num}_k \}$ can always be reliably identified with the accuracy corresponding to the chosen numerical scheme and the wave field discretization as illustrated in Fig. \ref{fig_3}(a).
Note that for a large particular eigenvalue $\zeta_k$ the accuracy $|\zeta^{num}_k - \zeta_k|$ is lower than for small-amplitude solitons.
As expected, the reduction of the domain makes the eigenvalues undetectable for solitons exposed to truncation, which is easy to observe in Fig. \ref{fig_3}(a) in comparison with the schematic wave field profile demonstrated in the bottom of Fig. \ref{fig_3}(c).

The main result of the work is presented in Figs. \ref{fig_3}(b,c) where the influence of $L$ on the errors of the calculated norming constants $\{ \rho^{num}_k \}$ is considered.
Firstly, we demonstrate the results of calculations when $\{ \zeta^{num}_k \}$ are computed from the condition (\ref{a_num}) with the standard double (machine) precision leading to $\delta \zeta_k \sim 10^{-16}$ in the expression (\ref{b_IST_N_lamba_k}).
Fig. \ref{fig_3}(b) shows that for large enough $L$ these deviations in eigenvalues lead to the exponential growth of errors as predicted by the developed theory, see Eq. (\ref{errors_rate}).
At the same time, for smaller $L < 15$, as expected, we observe the truncation errors for the norming constants.

These two effects make the whole set of solitons not possible to identify at any fixed value of $L$, in particular the third and fourth solitons shown by green and purple dots are completely ``uncatchable'' when standard precision is used.
Secondly, we perform the same set of simulations employing high-precision arithmetics to identify $\{ \zeta^{num}_k \}$ and calculate $\{ \rho^{num}_k \}$ excluding the described anomalous errors.
Fig. \ref{fig_3}(c) shows that we successfully obtain $\{ \rho^{num}_k \}$ when $L$ is large enough so that the wave field is well localised inside the computational domain.

{\it Conclusions}. -- 
In this work we consider $N$-soliton solutions of the nonlinear Schr\"odinger equation and the corresponding solutions of the Zakharov--Shabat problem.
The main result is the theoretical derivation of the connection between scattering coefficients $a_N$ and $b_N$ defined for a problem on the infinite line with $a_{N, tr}$ and $b_{N, tr}$ corresponding to the same problem on a finite (truncated) domain of the width $2L$.
Using the dressing method we obtain closed-form expressions for $a_{N, tr}$ and $b_{N, tr}$ allowing us to express $b_N = \lim b_{N, tr}$ at $L \to \infty$ and demonstrate its analytic properties.
Based on these results we reveal a new class of inherent instabilities of the direct scattering transform leading to {\it anomalous} numerical errors of norming constants  growing exponentially with $L$.
A high-precision arithmetic is required in order to exclude these errors leading to the fact that hybrid methods employing two different numerical approaches for a subsequent computing of eigenvalues and norming constants should be applied with caution due to possible systematic errors in eigenvalues.
Note that small variations of the input wave fields do not affect the overall stability of the DST.
We expect that the presence of solitons in complex wave fields containing continuous spectrum is always manifested by rapidly changing landscape of $\rho_{tr}(\zeta)$,
representing a general situation when the suggested strategy to perform the DST can be applied.
In particular this idea is supported by results for the exactly solvable rectangular potential model revealing exponential growth of the scattering coefficients in the complex plane with the potential length \cite{manakov1973nonlinear}.
In addition, our results can be straightforwardly generalised to another important
class of coherent structures appearing in the NLSE -- breathers \cite{kuznetsov1977solitons, ma1979perturbed, peregrine1983water, akhmediev1985generation, pelinovsky2008book}, 
as well as to other integrable models.
These insights give theoretical foundations to develop robust algorithms for the calculation of scattering data of complex wave fields to study various nonlinear phenomena.
The complete identification of coherent structures in stochastic wave fields is on top of the current agenda \cite{soto2016integrable, conforti2018auto,Randoux2018nonlinear, trillo2019quantitative, Kraych2019, gelash2019bound}.

{\it Acknowledgments}. -- 
Both authors (A.G. and R.M.) proposed key ideas and contributed equally to theoretical computations, numerical simulations and manuscript preparation.
A.G. acknowledges support of RFBR grant No. 19-31-60028, R.M. acknowledges support of RSF grant No. 19-79-30075.
Section ``Anomalous  errors'' reports the results of the work supported solely by RSF grant No. 19-79-30075.
The authors thank Profs. E.A. Kuznetsov, V.E. Zakharov and the group of Prof. D.A. Shapiro for fruitful discussions.


%

\end{document}